\newcommand{\CLASSINPUTtoptextmargin}{0.78in}
\newcommand{\CLASSINPUTbottomtextmargin}{1.1in}
\documentclass[conference]{IEEEtran}
\IEEEoverridecommandlockouts

\usepackage{cite}
\usepackage{amsmath,amssymb,amsfonts}
\usepackage{algorithmic}
\usepackage{graphicx}
\usepackage{textcomp}
\usepackage{xcolor}
\makeatletter 
\newcommand{\linebreakand}{%
  \end{@IEEEauthorhalign}
  \hfill\mbox{}\par
  \mbox{}\hfill\begin{@IEEEauthorhalign}
}
\makeatother 

\def\BibTeX{{\rm B\kern-.05em{\sc i\kern-.025em b}\kern-.08em
    T\kern-.1667em\lower.7ex\hbox{E}\kern-.125emX}}
\begin{document}

\title{Slotted Aloha  for  Optical Wireless Communications in Internet of Underwater Things}
\author{{Milica Petkovic,~Sotiris A. Tegos,  Panagiotis D.  Diamantoulakis, \\
Dejan Vukobratovic, Erdal Panayirci, \v Cedomir Stefanovi\' c, George K. Karagiannidis}
\thanks{This work has received funding from the European Union Horizon 2020 research and innovation programme under the grant agreement No 856967. This publication was based upon work from COST Action NEWFOCUS CA19111, supported by COST (European Cooperation in Science and Technology). This work was also supported by the Secretariat for Higher Education and Scientific Research of the Autonomous Province of Vojvodina through the project “Visible light technologies for indoor sensing, localization and communication in smart buildings” (142-451-2686/2021). }
\thanks{M. Petkovic and D. Vukobratovic are with Faculty of Technical Sciences, Novi Sad, Serbia (e-mails: \{milica.petkovic, dejanv\}@uns.ac.rs).}
\thanks{S. A. Tegos and P. D.  Diamantoulakis  are with Department of Electrical and Computer Engineering, Aristotle University of Thessaloniki, Greece (e-mails: \{tegosoti, padiaman\}@auth.gr).}
\thanks{E. Panayirci is with Kadir Has University, Cibali-Fatih,
Istanbul, Turkey
(e-mail:eepanay@khas.edu.tr).}
\thanks{\v C. Stefanovi\' c is with Aalborg University, Aalborg, Denmark 
(e-mail:cs@es.aau.dk).}
\thanks{G.  K. Karagiannidis is with Department of Electrical and Computer Engineering, Aristotle University of Thessaloniki, Greece, and with Cyber Security Systems and Applied AI Research Center, Lebanese American University (LAU), Lebanon (e-mails: geokarag@auth.gr).}
}

\maketitle

\begin{abstract}
In this work, we design and analyse a Slotted ALOHA (SA) solution for Optical Wireless Communication (OWC)-based Internet of Underwater Things (IoUT). In the proposed system, user devices  exchange data with an access point (AP) which exploits the capture effect. The space spanned by the IoUT nodes is three-dimensional, i.e., users are located in half-sphere centered at the AP placed  at the bottom of a floating object at the water surface level. The analytical expressions  for the  system throughput and reliability expressed in terms of the outage probability are derived. Based on the simulated signal-to-noise-and-interference-ratio statistics  and derived  analytical expressions, we present numerical results that investigate the trade-off between the system performance and the IoUT system parameters, such as the number of users, activation probability and type of water medium. The presented conclusions provide valuable insights into the design of an SA-based solution for IoUT communications.   
\end{abstract}


\section{Introduction}

According to \cite{ericsson}, the number of 
Internet of Things (IoT) connections are estimated  to grow from 13.2 billion in 2022 to 34.27 in 2028.
As a consequence, 5G and beyond-5G systems are faced with a challenge to guarantee  a reliable connectivity with high energy- and spectral-efficiency for extreme  IoT connection densities \cite{ericsson,6gvis,white}.
In the past decades, IoT systems have been extensively investigated, designed, and developed, pertaining to numerous applications in various domains, such as the smart cities \cite{r1}, smart home \cite{r2a,r2b}, healthcare \cite{r4a,r4b},  industry (including  smart transportation, smart energy,  smart factory) \cite{r3a,r3b}, and others \cite{IoT}. 

Since about 70$\%$ of the Earth surface is covered by water, the IoT applications have been extended to underwater environments, named Internet of Underwater Things (IoUT) \cite{UW11, UW22}.
The IoUT framework, mostly deployed as a part of the Smart Ocean \cite{SO}, represents a smart network of the intelligent, interconnected underwater objects, such as sensor nodes, autonomous underwater vehicles (AUV), boats, gliders, divers, and similar.
Unique characteristics of the underwater medium result in the different models and designs of IoUT compared to the classical land-based IoT systems. Underwater acoustic communications have been mostly deployed as wireless technology within IoUT \cite{acustic}, but low data rates and limited bandwidth of acoustic waves  are inadequate for supporting a large number of underwater devices while maintaining  requirements of the next-generation networks. 

As a promising solution to fulfill demands of the upcoming 6G systems,  optical wireless communications (OWC) have received attention as a potential technology for IoUT \cite{UW0}.
The OWC based underwater communications are convenient for IoUT application due to low power consumption, low latency, high bandwidth, low cost and security \cite{UW1,UW2,UW2a,UW3,UW3a,UW4}.
Due to a massive population of IoT devices, the connectivity for short and sporadic data transmissions has to be established in specific underwater environment. As IoUT transmission can be characterized by short packets and unpredictable device activity, random access  protocols, such as Slotted ALOHA (SA) \cite{Roberts}, have been often employed as a modern random access scheme in IoT system \cite{refRA,SA_NOMA}. To the best of our knowledge, the SA based OWC system in IoUT scenario  was only analyzed  in \cite{refMAC}.  Authors in \cite{refMAC} focused on the PHY/MAC cross-layer framework, while performing the analysis of the SA based OWC IoUT system performance, in terms of bit-error rate, success rate, delay and throughout.

The aim of the paper is to investigate the design an SA-based OWC system in IoUT scenario, and analyze the system reliability and throughput. The contributions of the paper are summarized as follows:
\begin{itemize}
  \item We present a novel design of future OWC-based IoUT system, considering three-dimensional (3D) network configuration. The system considered in \cite{refMAC} also observes 3D setup, with the IoUT devices  placed within a sphere centered at the access point (AP).
  Differently from \cite{refMAC}, we consider a 3D system with IoUT devices located within a half-sphere centered at the AP. In our system, OWC AP is fixed to the bottom of a floating object on the water surface level (see Fig. 1).
  \item We consider  uplink communication based  on SA with  capture effect, meaning that the OWC AP will try to decode the packet even if more than one user is active in a slot.  The main goal of the proposed setup is to investigate how SA with capture affects the system reliability and throughput in the underwater conditions.
  \item 
  We provide the the signal-to-interference-and-noise (SINR) analysis of proposed system taking into account specific underwater medium. The simulated SINR statistics is utilized to analyze  the overall system performance. 
  \item  We derive reception reliability of a randomly activated user in terms of outage probability, as well as the system throughput, by taking into account the interference contribution from other active users.
   \item  The derived expressions  are used to identify the trade-offs between the performance and the IoUT system parameters, which help  us to obtain valuable insights for the design of an SA-based solution for an OWC based IoUT system.  
\end{itemize}

The rest of the text is organized as follows.
Section II presents the system model, while in Section III we provide the performance analysis. Numerical results are given in Section IV. Section V concludes the paper.

\section{System Model for SA-based OWC IoUT Scenario}

\begin{figure}[b!]
\centerline{\includegraphics[width=3in]{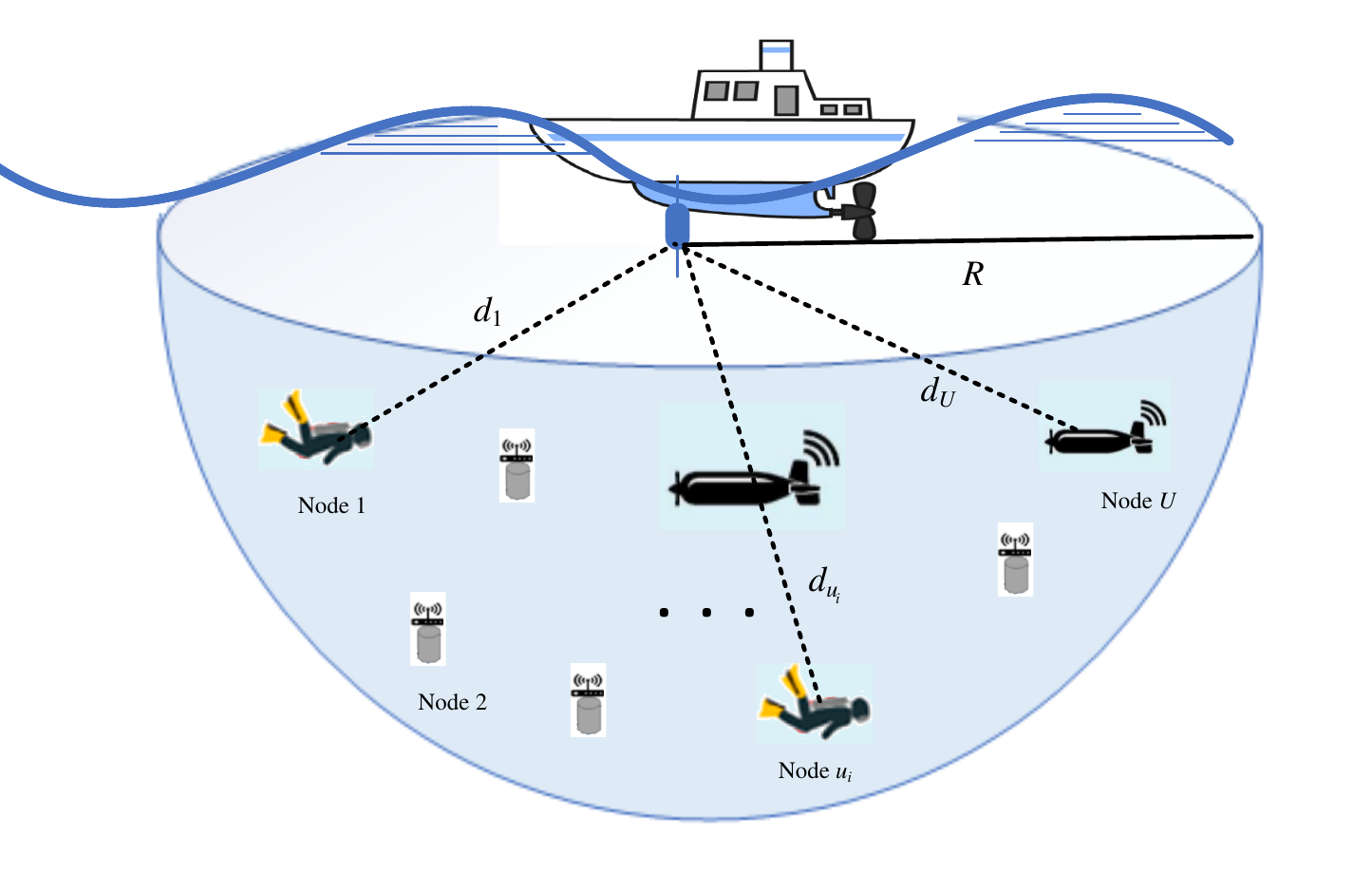}}
\caption{OWC-based IoUT system model.}
\label{Fig1}
\end{figure}

We analyze an uplink communication scenario in an OWC based IoUT framework. A total of $U$ OWC IoUT devices, equipped with OWC transmitters such as LEDs or laser diode, contend to access a common  AP. IoUT devices are uniformly placed within a half-sphere with radius $R$, centered at OWC AP, which is located at the bottom of a floating object (glider, buoy, AUV,  etc.) at the water surface  (see Fig.~\ref{Fig1}).

The SA protocol is adopted as a random access policy for the uplink transmission using slots that accommodate a single packet transmission. Every IoUT device is active with a certain probability $p_a$ in every slot (independently of the activity of other devices in the same slot).
A fixed-length packet fitting the slot is transmitted by a user if it is active in a corresponding slot.
We adopt notation where the set of active users in a slot is  denoted by $\mathcal{U}_a$, while $U_a = |\mathcal{U}_a|$ represents  number of active users with $U_a \in [0,U]$, which is a random variable distributed according to binomial distribution $\mathcal{B}(U, p_a)$.

The OWC-based IoUT devices employ LED-based sources that operate in  visible light or infra-red part of the spectrum. LED transmitters employ the intensity modulation  with non-return-to-zero on-off keying scheme, while the OWC photodetector receiver, representing the OWC AP, performs  direct detection (DD) of the received light intensity.
All IoUT devices transmit with the same optical power denoted by $P_t$.

In a slot with $U_a$ active users, the light intensity impinging the underwater OWC AP includes both  the contribution of $U_a$ active users and the background radiation noise. After optical-to-electrical conversion, with $\eta$ denoting the conversion coefficient, the received signal is
\begin{equation}
\begin{split}
y(t)= \sum_{i=1}^{U_a} P_t \eta h_i s_i(t) + n(t),
\end{split}
\label{y1}
\end{equation}
where $s_i(t)$, $i=1, \dots, U_a$, is the signal transmitted by the active user $u_i$, and  $h_i \geq 0$ represents the optical LoS gain from the $i$-th user to the AP. The Gaussian noise, comprised of background radiation and thermal noise and denoted by $n(t)$, is modeled as a zero-mean Gaussian random variable with variance $\sigma_n^2=N_0B$, where $N_0$ is the noise spectral density and $B$ is the system bandwidth.

\textit{SA with Capture Effect:} In a classical SA collision channel model, the transmitted packet will not be decoded if more than one user is active in a certain slot. One of the approaches to improve system performance of the SA scheme is to employ a receiver that exploits so called capture effect \cite{Roberts,Zorzi, IWOW, IoTjournal}, where the received packet may survive the collision and be correctly decoded even though more then one user is active in a corresponding slot. In our scenario, SA with capture model assumes that the underwater OWC AP receiver will attempt to decode the transmission despite the interference due to the other signals. For that reason, the 
received signal in (\ref{y1}) can be rewritten as
\begin{equation}
\begin{split}
y(t)= P_t \eta h_{1} s_{1}(t) + \sum_{i=2}^{U_a} P_t \eta h_i s_i(t) + n(t),
\end{split}
\label{y2}
\end{equation}
where $s_{1}(t)$ is the signal sent by a reference user and  $h_{1} \geq 0$ is the optical gain from the reference user to the AP. The interference contribution from all other active users except the reference one is determined by the summation term in (\ref{y2}).
The instantaneous SINR for the reference user can be defined as
\begin{equation}
\begin{split}
{\rm SINR}  = \frac{ P_t^2 \eta^2 h_{1}^2 }{ \sum_{i=2}^{U_a}P_t^2 \eta^2 h_i^2 + \sigma _n^2} 
= \frac{\gamma_{1}  }{ \gamma_{\rm I} + 1}
\label{sinr}
\end{split},
\end{equation}
where 
\begin{equation}
\gamma_{1}  = \frac{ P_t^2 \eta^2   h_{1}^2 }{\sigma _n^2},~\gamma_{\rm I}  = \sum_{i=2}^{U_a} \gamma_i,~\gamma_i  = \frac{ P_t^2 \eta^2 h_i^2}{\sigma _n^2}.
\label{gUa}
\end{equation}

\begin{figure}[t!]
\centerline{\includegraphics[width=3.4in]{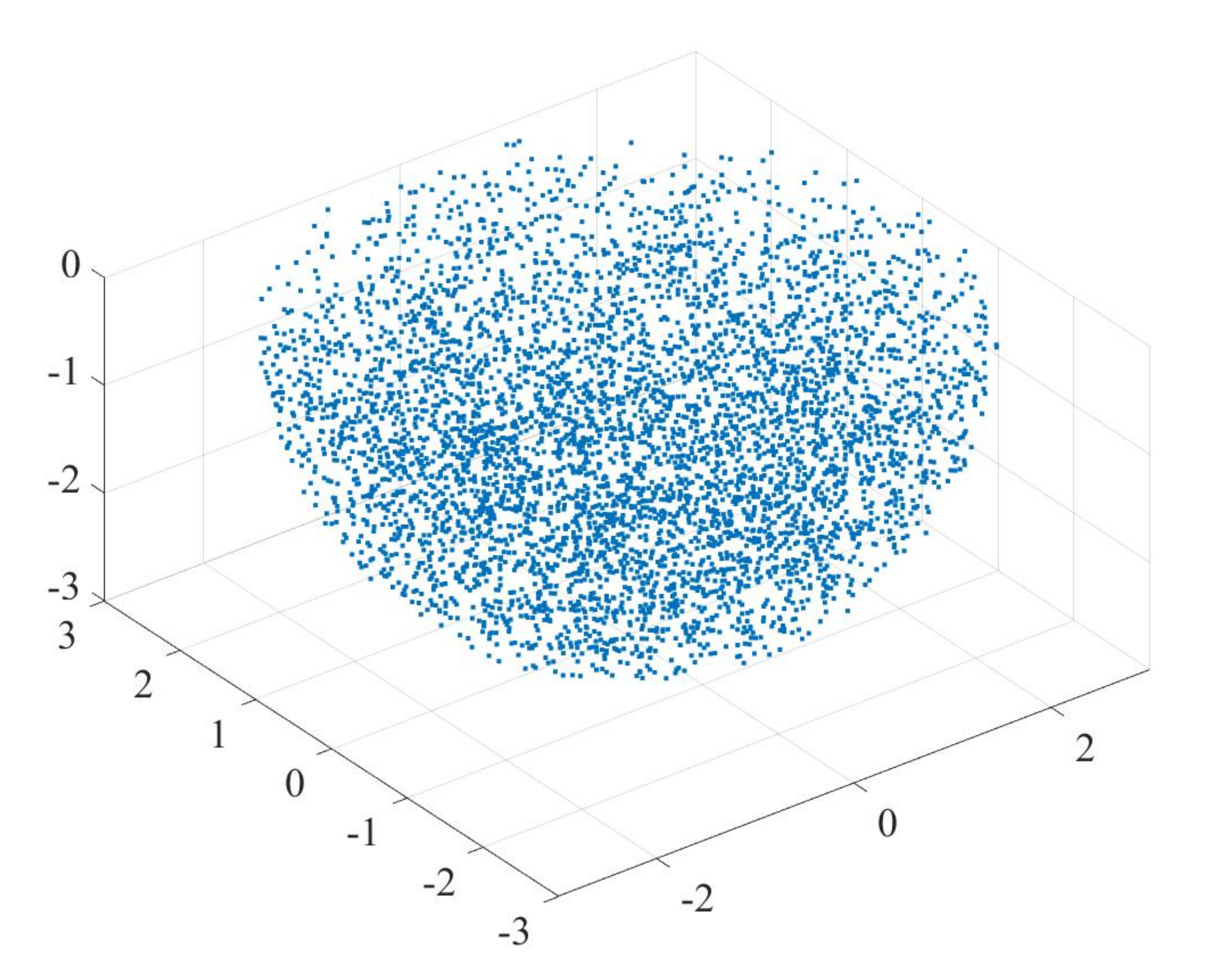}}
\caption{Uniformly distributed sampled points in the half-sphere, $R=3$ m.}
\label{Fig2}
\end{figure}

\textit{SINR Statistics for the OWC-based  IoUT:}
Propagation loss is determined by the channel attenuation due to absorption and scattering  described by Beer Lambert's law \cite{UW1,UW2, UW3, UW2a,UW3a,UW4}, resulting in the optical channel gain determined as 
\begin{equation}
h_i = \exp \left(-c(\lambda) d_i \right),
\label{hi}
\end{equation}
where $d_i$ is the Euclidean distance between the device $u_i$ and the OWC AP, while extinction coefficient $c(\lambda)$, dependent on the wavelength $\lambda $, represents the sum of the absorption and scattering coefficients
\begin{equation}
c(\lambda) = a(\lambda) + b(\lambda).
\label{c}
\end{equation}
Coefficients $c(\lambda)$, $a(\lambda)$ and $b(\lambda)$ are dependent on water types and depths (see Table I and \cite{UW1,UW2,UW3} for more details).

Recall that $U$ devices are assumed to be positioned within a half-sphere of radius $R$ (see Fig. 1). If the positions of users are sampled by a uniform distribution, we can generate \textit{N} samples of points distributed within the half-sphere of radius $R$ centred at the origin, i.e., the OWC AP. An example of simulated samples in MATLAB are presented in Fig. 2 considering $N=10^4$ and $R=3$ m.
From sampled positions of IoT devices, we can easily determined distances $d_i$ from all users and the AP, which can be further utilized to determine the optical channel gains $h_i$ based on (\ref{hi}). Finally, after defining values of $P_t$, $\eta$, $N_0$ and $B$, the samples  of $h_i $ are used to obtain the corresponding  values of SINR based on (\ref{sinr}), which helps us to numerically determine the PDF  ${f}_{{\rm SINR} } \left( \gamma | U_a \right)$ and the CDF $ {F}_{{\rm SINR} } \left( \gamma | U_a \right)$ of SINR, both dependent on the number of active users $U_a$. Simulated statistical characterization of the overall SINR is further used in our work to derive the outage probability and throughput of the considered underwater OWC system.

\section{Performance analysis: System reliability and Throughput}

\subsection{Outage probability}

The system reliability in terms of the outage probability can be determined by adopting the model that assumes the outage happens if the overall SINR falls below a predetermined threshold $\gamma_{\rm th}$.
For the system setup considered in this paper, by setting ${\rm SINR}=\gamma$ which is dependent on the number of active users $U_a$, we can calculate the   outage probability of the transmission from a randomly selected user (i.e., the reference user), as 
\begin{equation}
\begin{aligned}{ \rm P_{out}} ( U_a ) \! =\!\mathbb{P}\left[{ \gamma \!<\! \gamma_{\rm th}} |  U_a \right]\! = \! F_{{\rm SINR}}\left( \gamma_{\rm th}| U_a \right).
\end{aligned}
\label{Pout}
\end{equation}
The unconditional outage probability can be determined as 
\begin{align}
    {\rm P_{out}} = \sum_{k = 1}^{U} { \rm P_{out}} (  U_a = k  ) \, \mathbb{P} [ U_a = k ],
    \label{Pout1}
\end{align}
where $\mathbb{P} [ U_a = k ]$ for a Bernoulli arrival process is defined as
\begin{equation}
\label{binom}
\mathbb{P} [ U_a = k ]  = \binom{U}{k} p^k ( 1 - p)^{U-k}, \;~~ k = 0, \dots, U.
\end{equation}
The system reliability can be  calculated as ${ \rm P_R} = 1 - {\rm P_{out}}$.

\subsection{Throughput for SA}

The  system throughput considering the effects of the noise and the interference among randomly activated users  can be defined as 
\begin{equation}
\begin{split}
    T & = 0 \cdot \mathbb{P} [ U_a = 0] + \sum_{k=1}^{U}  (  1 -  {\rm P_{out}} (  U_a = k  ) ) \mathbb{P}  [ U_a = k ]  \\
    & =  ( \mathbb{P} [U_a > 0 ]  -  {\rm P_{out}}  =  ( 1-(1-p_a)^U  -  {\rm P_{out}}),
\end{split}
\label{T}
\end{equation}
where $\mathbb{P} [ U_a > 0] = 1-(1-p_a)^U$ defines the probability that at least one user is active during the slot.

\begin{table}[!t]
\caption{Values of the contribution underwater absorption,
scattering, and extinction coefficients \cite{UW1,UW2,UW3}.}
\begin{center}
\begin{tabular}{|c|c|c|c|}
\hline
~  & $a(\lambda)~{\rm in~ m}^{-1}$& $b(\lambda)~{\rm in~ m}^{-1}$ &$c(\lambda)~{\rm in~ m}^{-1}$ \\
\hline
Pure sea water & 0.053 & 0.003 & 0.056 \\
\hline
Clear ocean water & 0.069 & 0.08 & 0.15 \\
\hline
Coastal ocean water & 0.088 & 0.216 & 0.305 \\
\hline
Turbid harbor water & 0.295 & 1.875 & 2.17 \\
\hline
\end{tabular}
\label{tab1}
\end{center}
\end{table}

\section{Numerical results and Discussion}

\begin{figure}[!t]
\centerline{\includegraphics[width=3in]{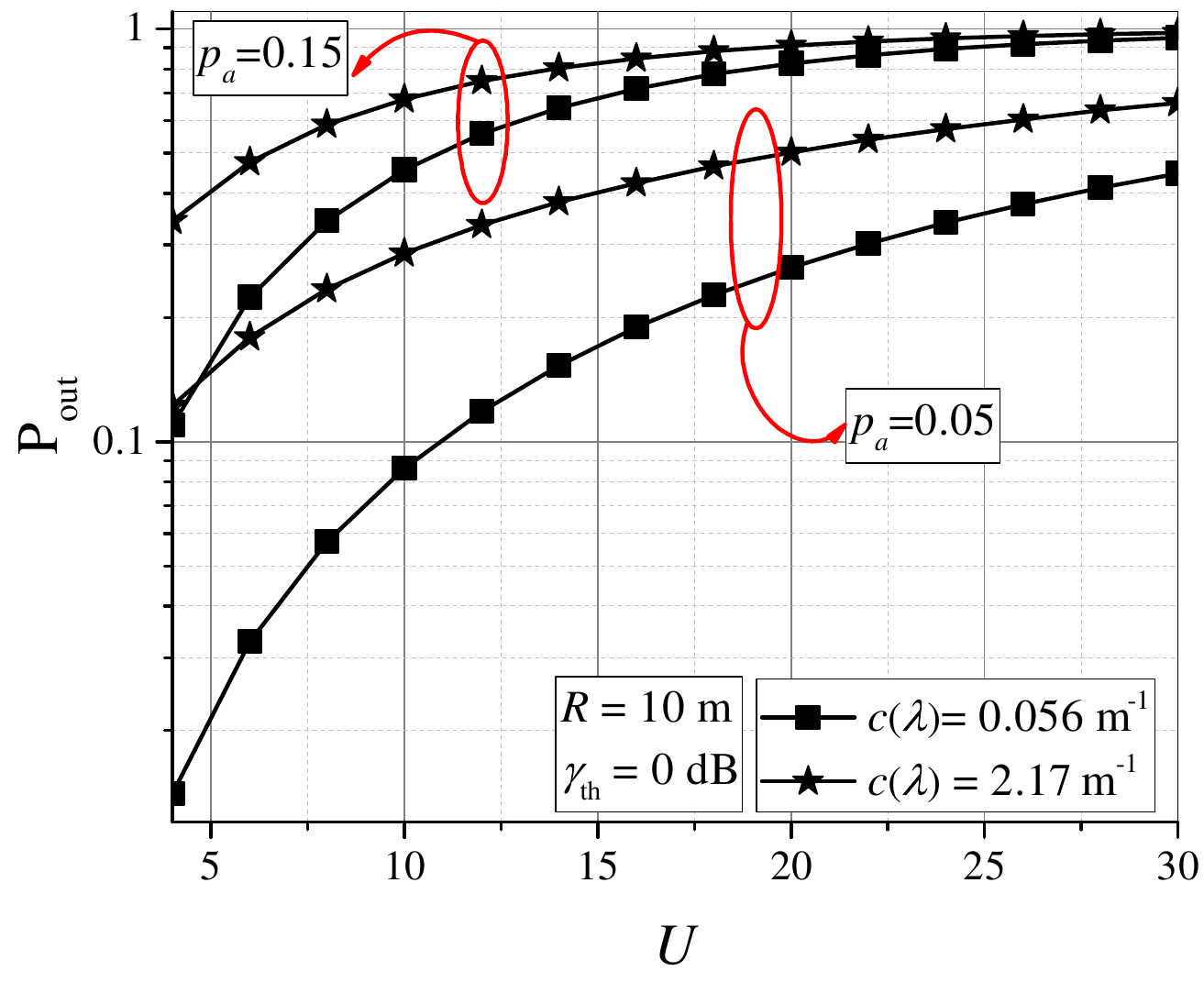}}
\vspace{-0.1in}
\caption{Outage probability vs. $U$ for different values of  coefficient $c(\lambda)$.}
\label{Fig3_Pout}
\vspace{-0.1in}
\end{figure}

In this section, we present numerical results obtained by Monte Carlo simulations and derived expressions (\ref{Pout1}) and (\ref{T}) to provide  basic insights for the OWC IoUT system design. The proposed scenario is analyzed by adopting the following values for the  parameters: the conversion efficiency is $\eta=0.8$, while the noise power spectral density is $N_0=10^{-21}~{\rm W}/{\rm Hz}$, and the system bandwidth is $B=200~{\rm kHz}$. The OWC capture threshold is $\gamma_{\rm th}=0$ dB and transmitted optical power is $P_t = 100$ mW. The values of the coefficient related to underwater channel conditions are given in Table I for different types of water.

Fig.~\ref{Fig3_Pout} presents the outage probability dependence on the total number of IoUT devices located within the half-sphere of radius $R=10$ m. Besides different water environment (pure sea and turbid harbor water), two values of activation probabilities are considered. Higher activation probability reflects in greater number of active users in a slot, which results in  stronger interference contribution and greater probability of outage. Hence, higher $p_a$ will lead to the deterioration of the system performance. As it is expected, higher $c(\lambda)$, meaning that turbidity in water is stronger, will cause the worse system performance. From  Fig.~\ref{Fig3_Pout} we can conclude that the effect of the type of water medium has stronger effect on  ${\rm P_{out}}$ when the number of (active) users is lower, i.e., lower values of $p_a$ and $U$.

\begin{figure}[!t]
\centerline{\includegraphics[width=3in]{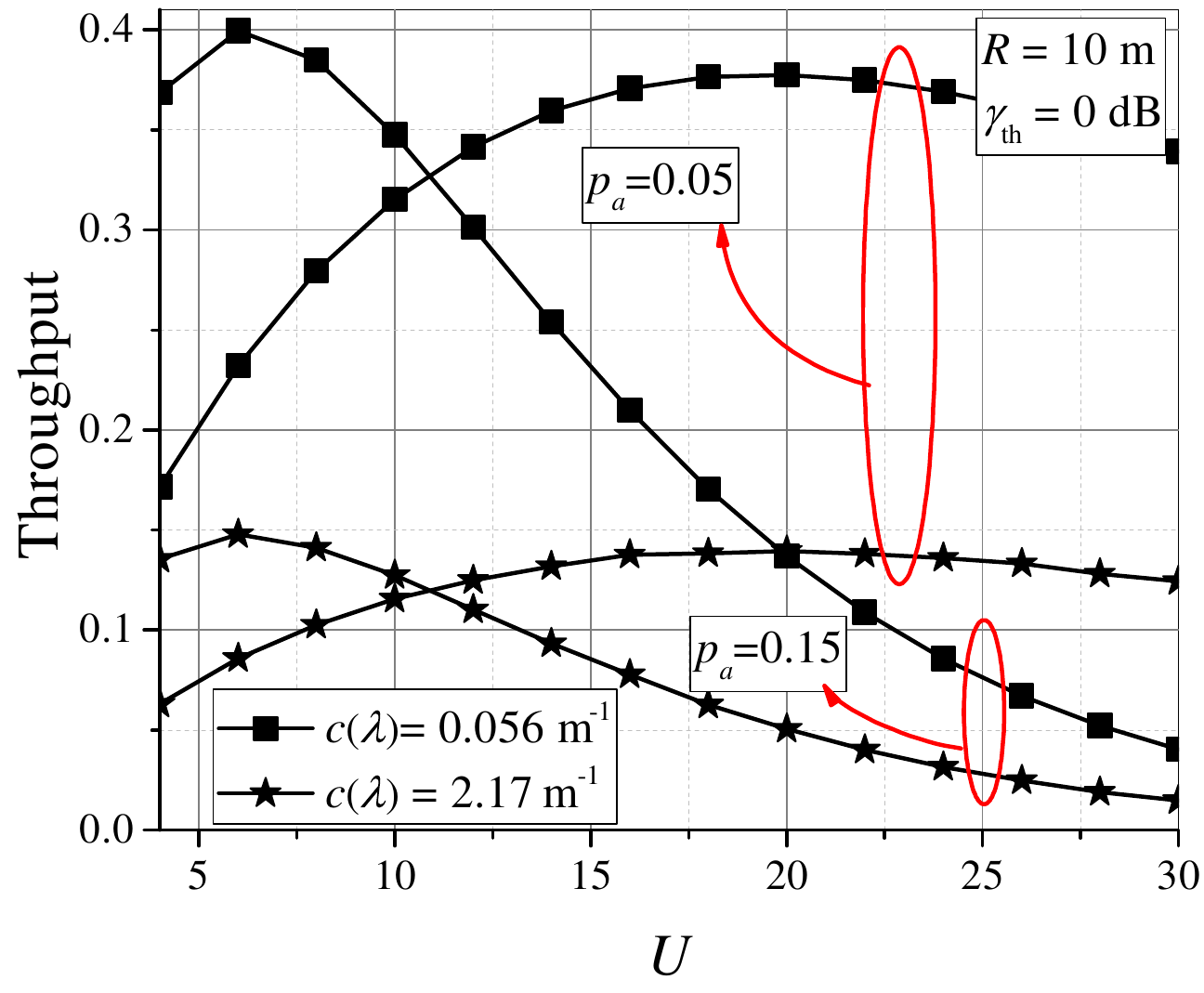}}
\vspace{-0.1in}
\caption{Throughput vs. $U$ for different values of underwater coefficient $c(\lambda)$.}
\label{Fig3_T}
\vspace{-0.1in}
\end{figure}

\begin{figure}[!t]
\centerline{\includegraphics[width=3in]{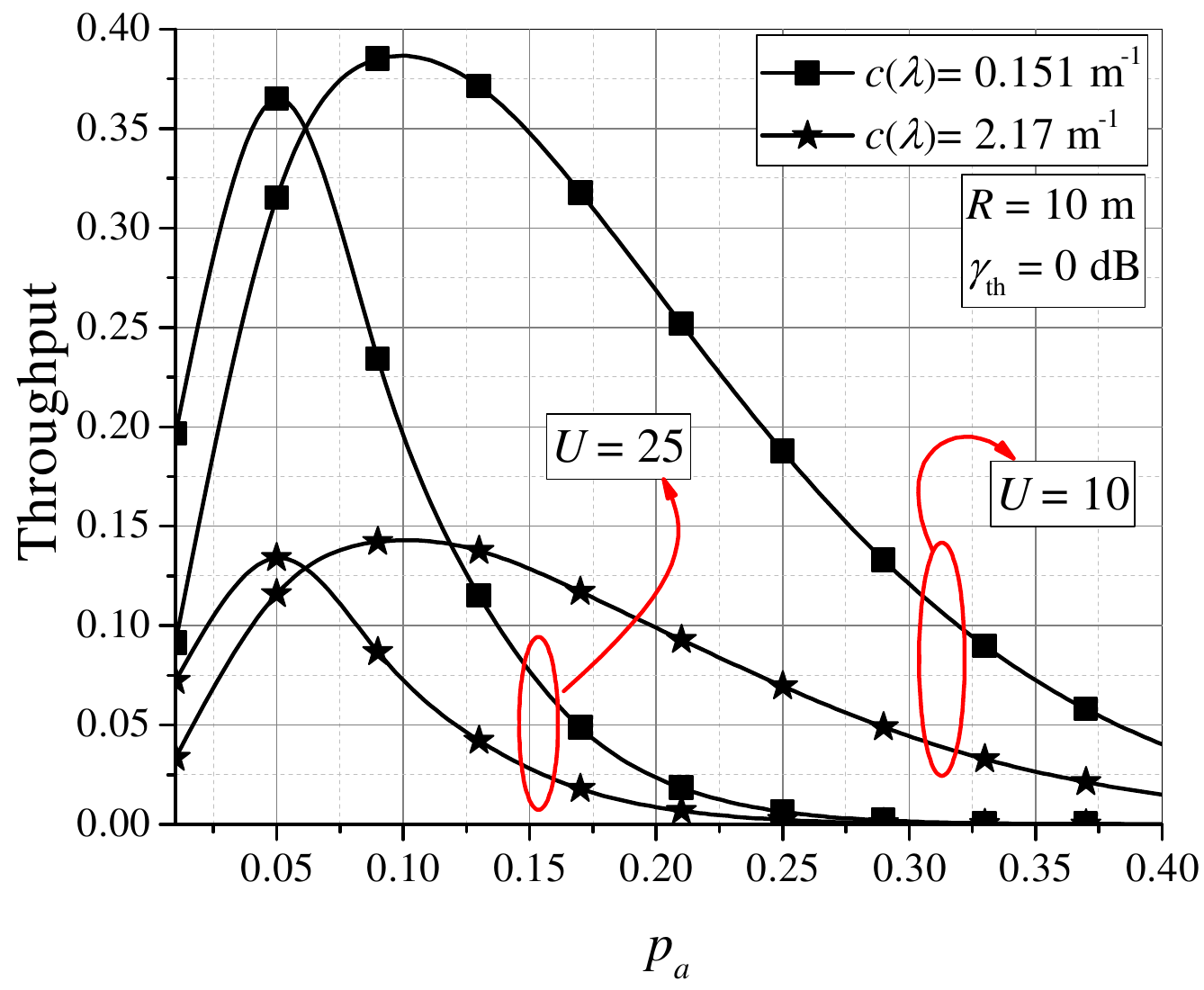}}
\vspace{-0.1in}
\caption{Throughput vs. $p_a$  for different values of underwater coefficient $c(\lambda)$.}
\label{Fig1_T}
\vspace{-0.1in}
\end{figure}
Throughput dependence on the  total number of IoUT devices $U$ is shown in  Fig.~\ref{Fig3_T}, for different values of $p_a$ in pure sea and turbid harbor water mediums. In contrary to previous figure and the corresponding discussion for outage probability performance, throughput does not show monotone behavior when the number of total users increases. For example, for $c(\lambda)$=0.056 m$^{-1}$, throughput is higher when $p_a=0.15$ and $U<11$. On the other hand, lower activation probability ($p_a=0.05$) reflects in better throughput  for $U>11$. It is obvious that the maximal value of throughput exists for a certain number of active users, which will be further discussed next.

\begin{figure}[!t]
\centerline{\includegraphics[width=3in]{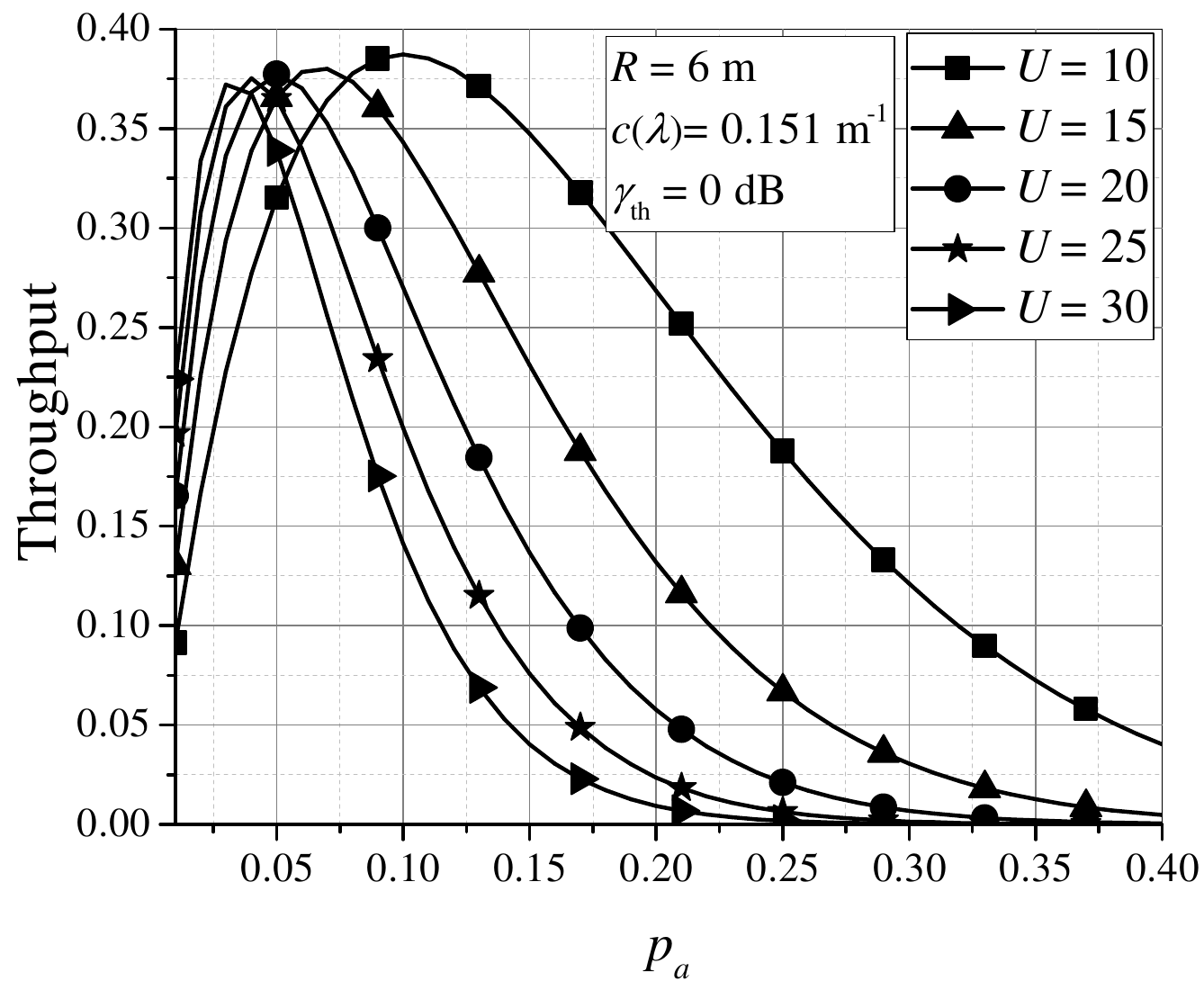}}
\vspace{-0.1in}
\caption{Throughput vs. $p_a$ for different number of IoT users $U$.}
\label{Fig2_T}
\vspace{-0.2in}
\end{figure}

Throughput dependence on $p_a$ is presented in Fig.~\ref{Fig1_T} for different values of the total number of IoUT devices $U$ in different water media. 
As previously stated, higher  $p_a$ and  higher $U$ imply more active users in a slot, which determines the overall SINR and has a direct effect  on the potential of the capture.  It can be observed that maximal value of throughput exists for an optimal value of $p_a$. After the maximum value  is reached, the system throughput starts decreasing with a further lowering of $p_a$. The optimal value of $p_a$ remains the same for different types of water conditions, but maximal value of throughput (as well as the overall throughput performance) will increase with less turbid water (lower $c(\lambda)$). On the other hand, the optimal value of $p_a$ changes with different total number of users $U$, which can be also seen in Fig.~\ref{Fig2_T}. When the total number of users $U$ in the system is higher, the optimal value of $p_a$ that maximizes $T$ gets lower. 

Finally, we can conclude that  the total number of underwater IoUT devices and the activation probability  have crucial impact on the optimal system performance. Both parameters should be taken into consideration during the design of the access protocol.
For example, in a practical system scenario, the OWC AP can perform an estimation of the total number of users located in some half-sphere area below. Based on this estimate, the AP can decide which value of activation probability $p_a$ will result in maximal throughput and forward that information to  all the users of the IoUT system.

\section{Conclusion and Future Work}

In this paper, we presented a design and analysis of the uplink of an OWC-based IoUT system. The considered setup is a novel 3D IoUT system model, where users are located within a half-sphere centered at the bottom of a floating object placed at the water surface.
The proposed system employs a SA-based access scheme with the capture effect. 
Based on the simulated SINR statistics and derived  analytical expressions for the outage probability and the overall throughput, numerical results are obtained and   further analyzed to assess the system performance behavior depending on the system and channel parameters.

The presented numerical study suggests that the maximal throughput appears for a specific system load, i.e., the number of active users in a slot. Thus, the number of underwater IoT devices and  value of $p_a$ have an impact on the optimal system performance and should be  taken into consideration during the design of the access protocol. Next steps of our work include the mathematical derivation of SINR statistics for 3D IoUT network, as well as additional effects of water medium attenuation, such as geometric loss and turbulence effect.



\end{document}